\documentclass{PoS}
\let\ifpdf\relax
\pdfoutput=1
\title{Viewing the Chemical Evolution of the Quark-Gluon Plasma with Charge Balance Functions}

\ShortTitle{Chemical Evolution of the QGP}

\author{\speaker{Scott Pratt}
\thanks{The author is indebted to Claudia Ratti for providing lattice data, to Hui Wang for providing preliminary STAR data, and to Gary Westfall for providing the efficiency routine to simulate STAR's acceptance. This work was supported by the National Science Foundation's Cyber-Enabled Discovery and Innovation Program through grant NSF-0941373 and by the Department of Energy Office of Science through grant number DE-FG02-03ER41259.}\\
        Department of Physics and Astronomy\\
        and National Superconducting Cyclotron Laboratory\\
        Michigan State University, East Lansing, MI 48824\\
        E-mail: \email{prattsc@msu.edu}}

\abstract{
Correlations from charge conservation are affected by when charge/anticharge pairs are created during the course of a relativistic heavy ion collision. For charges created early, balancing charges are typically separated by the order of one unit of spatial rapidity by the end of the collision, whereas those charges produced later in the collision are far more correlated. By analyzing correlations from STAR for different species, I show that one can distinguish the two separate waves of charge creation expected in a high-energy collision, one at early times when the QGP is formed and a second at hadronization. Further, I extract the density of up, down and strange quarks at in the QGP and find agreement at the 20\% level with expectations for a chemically thermalized plasma.
}

\FullConference{Critical Point and the Onset of Deconfinement\\
		March 11-15, 2013\\
		Napa, California, USA}

\begin{document}

\section{Introduction}
Charge is locally conserved. For every up, down or strange quark produced in a heavy ion collision, an anti-up, anti-down or anti-strange quark is produced at the same point in space time. In the canonical picture of the quark-gluon plasma (QGP), the system thermally equilibrates during the first fm/$c$, which requires several hundred pairs of quark-antiquark pairs to be created per unit rapidity in a central collision at RHIC (Relativistic Heavy Ion Collider). In an isentropic expansion, assuming the degrees of freedom are massless, the number of quarks would stay fixed until hadronization. During that time, the balancing charges drift apart. The average separation along the beam axis depends on the diffusion constant, the initial formation time, and the formation mechanism, and might become near or perhaps larger than one unit of spatial rapidity. At hadronization, a second wave of quark production occurs \cite{Bass:2000az}. Entropy conservation suggests that for every quark in the QGP, one would expect approximately 1.2 hadrons in the hadronic state, and each hadron has multiple quarks. Simply counting the number of quarks in the hadron gas suggests that the number of up and down quarks would nearly triple, while the number of strange quarks would remain approximately the same.

\begin{figure}
\centerline{\includegraphics[width=0.5\textwidth]{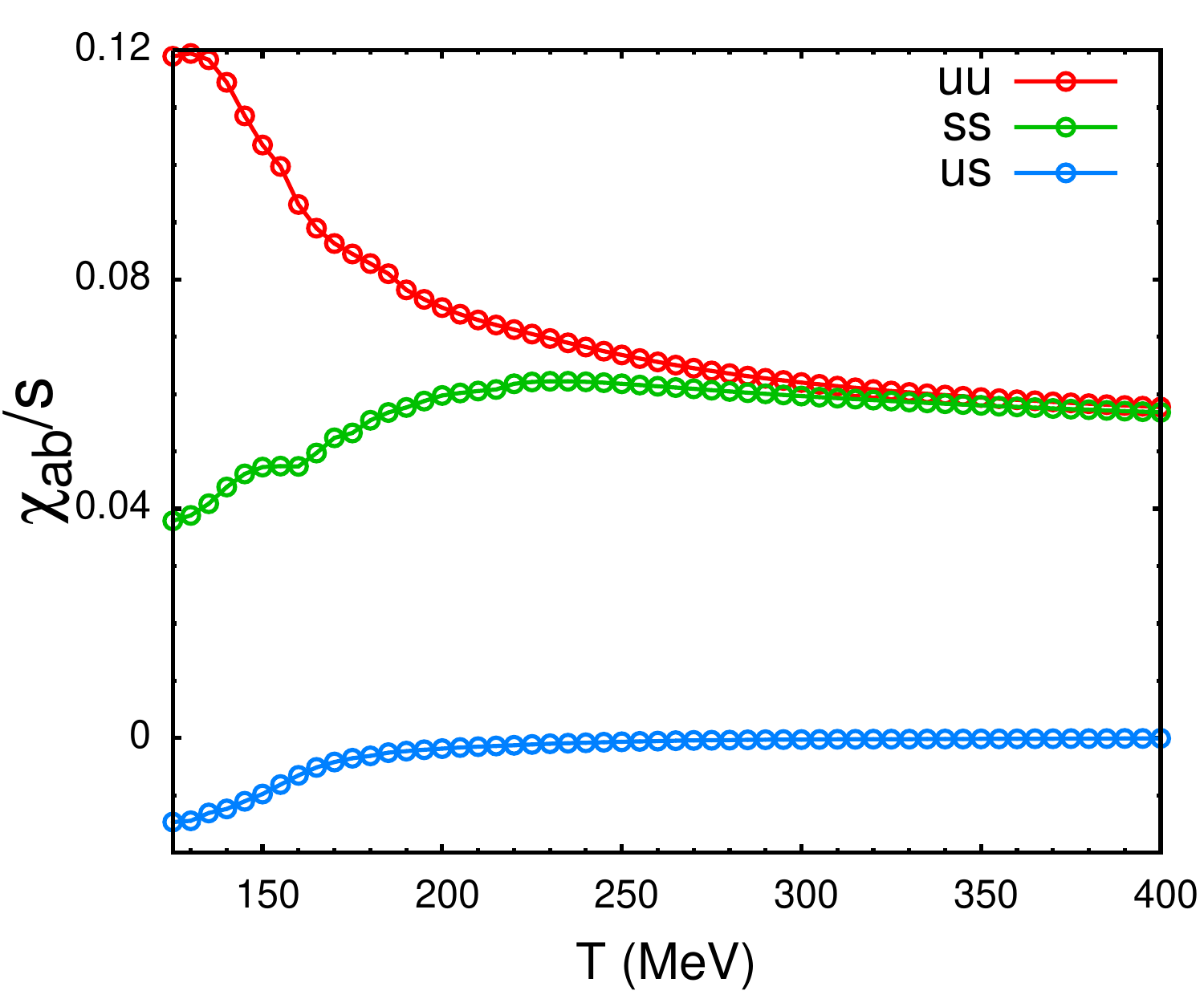}}
\caption{\label{fig:claudia}
Charge fluctuations as calculated by lattice gauge theory \cite{Borsanyi:2011sw,Ratti:2011au}. Results for temperatures below 160 MeV are consistent with a hadron gas (lattice data provided by C. Ratti). For a fixed entropy, there are increased numbers of up and down quarks in the hadronic phase, whereas the number of strange quarks is slightly smaller. The off-diagonal $us$ element comes from kaons, which contain a quark of each flavor. The off-diagonal element disappears above $T_c$ when the kaons dissolve. 
}
\end{figure}

The simple situation outlined above is based on the QGP being formed of massless quarks and gluons with the ratio of the quark number to entropy being constant during the QGP pahse. After the prolonged QGP phase, hadronization occurs in a narrow window. Lattice calculations \cite{Borsanyi:2011sw,Ratti:2011au} show that this picture is approximately true. Figure \ref{fig:claudia} shows that charge fluctuations, defined as
\begin{equation}
\chi_{ab}\equiv\frac{\langle Q_aQ_b\rangle}{V},
\end{equation}
stay approximately constant throughout the QGP phase after being scaled by the entropy density. Here, $Q_a$ is the net up, down or strange charge within a volume $V$. We are assuming throughout this talk that there is no net charge, $\langle Q_a\rangle=0$, which is true for lattice analyses, and is a good approximation for mid-rapidity measurements at the LHC or at the highest RHIC energies. For a parton gas,
\begin{equation}
\chi_{ab}^{\rm(qgp)}=\delta_{ab}\left(n_a+n_{\bar{a}}\right),
\end{equation}
where $n_a$ and $n_{\bar{a}}$ are the densities of up and down quarks. This follows because for a non-interacting gas, the only correlations are those between a particle and itself. For a non-interacting hadron gas the dominant correlations are those between quarks within the same hadron, and
\begin{equation}
\chi_{ab}^{\rm(had)}=\sum_\alpha n_\alpha q_{\alpha,a}q_{\alpha,b},
\end{equation}
where $q_{\alpha,a}$ is the charge of type $a$ of a hadron where the species is labeled by $\alpha$. The approximate constancy of $\chi_{ab}/s$ in Fig. \ref{fig:claudia} suggests that the density of quarks remains roughly constant within an expanding and co-moving hydrodynamic volume element. The rise of the ratio at low temperature is driven by the rise of the total number of quarks in the hadronic stage. Further, the off-diagonal element $\chi_{us}$ is indicative of the existence of hadrons \cite{Koch:2005vg}. 

The purpose of this letter is to show how recent STAR measurements of balance functions can be used to validate this two-wave nature of charge formation, and to extract the ratio $\chi_{ab}/s$ in the QGP phase. To do this, we must overcome the obvious constraint, that unlike lattice calculations, the charge does not fluctuate when integrated over the entire collision volume, and that one must look through the haze of hadronization. This will be accomplished by focusing on charge correlations, which describe the probability for having two charges separated in spatial rapidity by $\Delta\eta$,
\begin{equation}
g_{ab}(\Delta\eta)=\langle \rho_a(0)\rho_b(\Delta\eta)\rangle,
\end{equation}
where $\rho_a$ is the charge density of type $a$, and could be a density per unit volume or per unit spatial rapidity, depending on the context. Even though $\langle\rho_a\rangle=0$, the correlation $g_{ab}$ can differ from zero due to correlations. Further below, we show how $g_{ab}$ can be used to extract $\chi_{ab}$.

Charge conservation demands
\begin{equation}
\int d\Delta\eta~g_{ab}(\Delta\eta)=0.
\end{equation}
Assuming that $\chi_{ab}$ in the grand-canonical treatment is driven solely by short range correlations, one can state that the correlation for times before hadronization should have a form similar to
\begin{equation}
g_{ab}(\Delta\eta,\tau<\tau_{\rm(had)})=\chi_{ab}^{\rm(qgp)}\left\{
\delta(\Delta\eta)-\frac{e^{-(\Delta\eta)^2/2\sigma^2_{\rm (qgp)}}}{(2\pi\sigma^2_{\rm(qgp)})}
\right\}.
\end{equation}
More generally, the Gaussian form could be replaced by any function that integrates to unity, but one would expect a Gaussian form if quarks appeared suddenly, at a single time, followed by diffusion \cite{Pratt:2012dz}. If partons are created over a finite time interval, the single wave could be replaced by an integral over time with each time element contributing proportional to $\delta t(d\chi/dt)$, and with Gaussian widths corresponding to the diffusion width corresponding to the transport from that time element. The initial formation mechanism might also affect the form. Quark-antiquark pairs created by the tunneling process associated with breaking of longitudinal flux tubes could be pulled apart by the tunneling process. Thus, it would not be surprising if a more physical form for the correlation would have a longer non-Gaussian tail. Nonetheless, for this study, we consider a simple Gaussian form. 

Immediately after hadronization, the correlation of a particle with itself is defined by $\chi^{\rm(had)}$. The long-range correlation cannot change suddenly, because the newly created pairs are uncorrelated with charge far away. Thus to satisfy the condition that $g_{ab}$ integrates to zero, the correlation could have a form similar to
\begin{equation}
g_{ab}(\Delta\eta,\tau>\tau_{\rm(had)})=\chi_{ab}^{\rm(had)}\delta(\Delta\eta)
-(\chi_{a}b^{\rm(had)}-\chi_{ab}^{\rm(qgp)})\frac{e^{-(\Delta\eta)^2/2\sigma^2_{\rm (had)}}}{(2\pi\sigma^2_{\rm(had)})}
-\chi_{ab}^{\rm(qgp)}\frac{e^{-(\Delta\eta)^2/2\sigma^2_{\rm (qgp)}}}{(2\pi\sigma^2_{\rm(qgp)})}.
\end{equation}
Here, the width $\sigma_{\rm(had)}$ would be narrow, and in the limit of sudden hadronization one could justify the Gaussian form if the spread were diffusive.

If one measured all hadrons immediately after hadronization, $g_{ab}$ could be reconstructed from the correlations between hadrons. Unfortunately, neutrons are never measured, and weak decays make it impossible to fully reconstruct the strangeness. Thus, we consider correlations of hadrons,
\begin{equation}
G_{\alpha\beta}(\Delta\eta)=\left\langle [N_\alpha(\eta)-N_{\bar{\alpha}}(\eta)][N_\beta(\eta+\Delta\eta)-N_{\bar{\beta}}(\eta+\Delta\eta)]\right\rangle,
\end{equation}
where $\alpha$ and $\beta$ refer to specific hadronic species. In reference \cite{Pratt:2012dz} it was shown that $G_{\alpha\beta}(\Delta\eta)$ could be extracted from $g_{ab}$ by assuming additional charge is spread amongst the various hadrons thermally, i.e. to maximize entropy. The expression is
\begin{eqnarray}
G_{\alpha\beta}(\eta)&=&4\sum_{abcd}\langle n_{\alpha}\rangle
q_{\alpha,a}\chi^{{\rm(had)}(-1)}_{ab}(0)g^{{\rm(had)}}_{bc}(\eta)
\chi^{{\rm(had)}(-1)}_{cd}(\eta)
q_{\beta,d}\langle n_{\beta}\rangle.
\end{eqnarray}
Finally, one must account for the thermal smearing in mapping from spatial rapidity, $\eta$, to momentum rapidity, $y$. As explained in \cite{Pratt:2012dz}, this can be done by convoluting the correlations $G_{\alpha\beta}$ with a blast wave model. For this smearing, two parameters come into play: the final-state temperature and radial collective flow velocity. These parameters are taken to fit the mean $p_t$ of pions and protons reported by STAR. The collective velocity is $u_\perp=0.732c$, and the temperature is 102 MeV. Additionally, the decays of emitted particles are also taken into account by mapping the pre-decay correlations onto post-decay correlations by simulating decays.

Only those resonances stable to strong decays are considered for the calculations shown here. The hadron multiplicities are taken from assuming that particles were created according to chemical equilibrium with a temperature of 170 MeV. After accounting for all strong decays, the final state multiplicities of hadrons was determined. Analyses of RHIC data suggest this should provide a reasonable representation of the multiplicities at RHIC if one additionally accounts for baryon annihilation in the hadronic phase. We thus apply an additional baryon annihilation factor, $B_{\rm reduction}$. The adjustable model parameters are summarized in Table \ref{table:parameters}.
\begin{table}
\begin{tabular}{|c|c|c|c|}
\hline
parameter & description & expectation & MCMC range\\
\hline
$\sigma_{\rm qgp}$ & \parbox{3.0in}{\baselineskip=12pt
\vspace*{3pt}Spread of balancing charges created in initial thermalization of QGP\vspace*{3pt}}
& $0.6 - 1.2$ & $0.3 - 1.5$\\
$\sigma_{\rm had}/\sigma_{\rm qgp}$ &\parbox{3.0in}{\baselineskip=12pt
\vspace*{3pt}
Ratio of spread of balancing charges created at hadronization to the spread of those charges created at thermalization\vspace*{3pt}} & $0.2 - 0.4$ & $0 - 1$\\
$N_q/N_h$ & \parbox{3.0in}{\baselineskip=12pt
\vspace*{3pt}
Ratio of quarks in QGP to final state hadrons\vspace*{3pt}} & $\sim 0.85$ & 0.5 - 1.5\\
$s /u = s/d$ &\parbox{3.0in}{\baselineskip=12pt
\vspace*{3pt}
Ratio of strange quarks to up or down quarks\vspace*{3pt}}& $0.9 - 0.95$ & 0 - 1\\
$B_{\rm reduction}$ &\parbox{3.0in}{
\baselineskip=12pt
\vspace*{3pt}
Sets baryon yield relative to thermalized yield\vspace*{3pt}}& $\sim 2/3$ & $0.6-0.8$\\
\hline
\end{tabular}
\caption{\label{table:parameters}
Parameters varied in fitting experimental data. Expected values and the ranges varied through the MCMC trace are shown in the last two columns.
}
\end{table}

Calculations were performed using the models presented in \cite{Pratt:2012dz} and compared to preliminary balance functions from STAR \cite{Wang:2012jua}. Balance functions, $B_{\alpha\beta}$, are related to the correlations $G_{\alpha\beta}$, by a factor of the multiplicity,
\begin{equation}
B_{\alpha\beta}(\Delta\eta)=\frac{G_{\alpha\beta}(\Delta\eta)}{n_\beta + n_{\bar{\beta}}},
\end{equation}
where $n_\beta$ is the number per rapidity of the hadron species $\beta$.

In \cite{Pratt:2012dz}, it was shown that comparing $\pi^+\pi^-$, $K^+K^-$ and $p\bar{p}$ balance functions allows one to distinguish both scales, $\sigma_{\rm(qgp)}$ and $\sigma_{\rm(had)}$. Whereas the width of the $\pi^+\pi^-$ balance function was mainly determined by $\sigma_{\rm(had)}$, $\sigma_{\rm(qgp)}$ was the main factor for determining the width of the $p\bar{p}$ balance function. Both balance functions were affected by the up/down content of the quark gluon plasma. Not surprisingly, the $K^+K^-$ balance function was found to be sensitive to the strangeness content of the QGP. If the QGP achieves chemical equilibrium with regards to strangeness, few strange quark pairs are created during hadronization and the width of the $K^+K^-$ balance function is driven by the larger of the two widths. If the strangeness content of the QGP were half the equilibrium value, the narrower component of the correlation plays a strong role and the balance function is much narrower.

Our goal is to determine whether STAR data constrains the first four parameters from table I: the two diffusive widths,  the number of up/down quarks, and the number of strange quarks in the QGP. Since the balance functions, especially the $p\bar{p}$ balance function, are known to be sensitive to the final-state baryon yields, and since baryon yields are somewhat uncertain, a fifth parameter was introduced to account for annihilations. Hadronic yields were taken from the thermal yields taken from a hydrodynamic model assuming chemical equilibrium occurred at $T=170$ MeV \cite{Pratt:2012dz}, and assuming that the only chemical reactions thereafter were from strong-interaction decays and fusions. Baryon annihilations were not included in the thermal yields, but one would expect annihilations to reduce the baryon yields by approximately one third \cite{Pan:2012ne,Steinheimer:2012rd}, an expectation which has some experimental support \cite{Adler:2003cb}. To incorporate both annihilation processes and the uncertainty in the yields, baryonic yields were reduced by a factor $B_{\rm reduction}$, which was allowed to vary in the calculations here from 0.6 to 0.8. Finally, the correlations were superimposed onto a model of the STAR acceptance and efficiency. Unfortunately, our current efficiency filter in incomplete, so we fit only the shape of the balance function.

\begin{figure}
\centerline{
\includegraphics[width=0.44\textwidth]{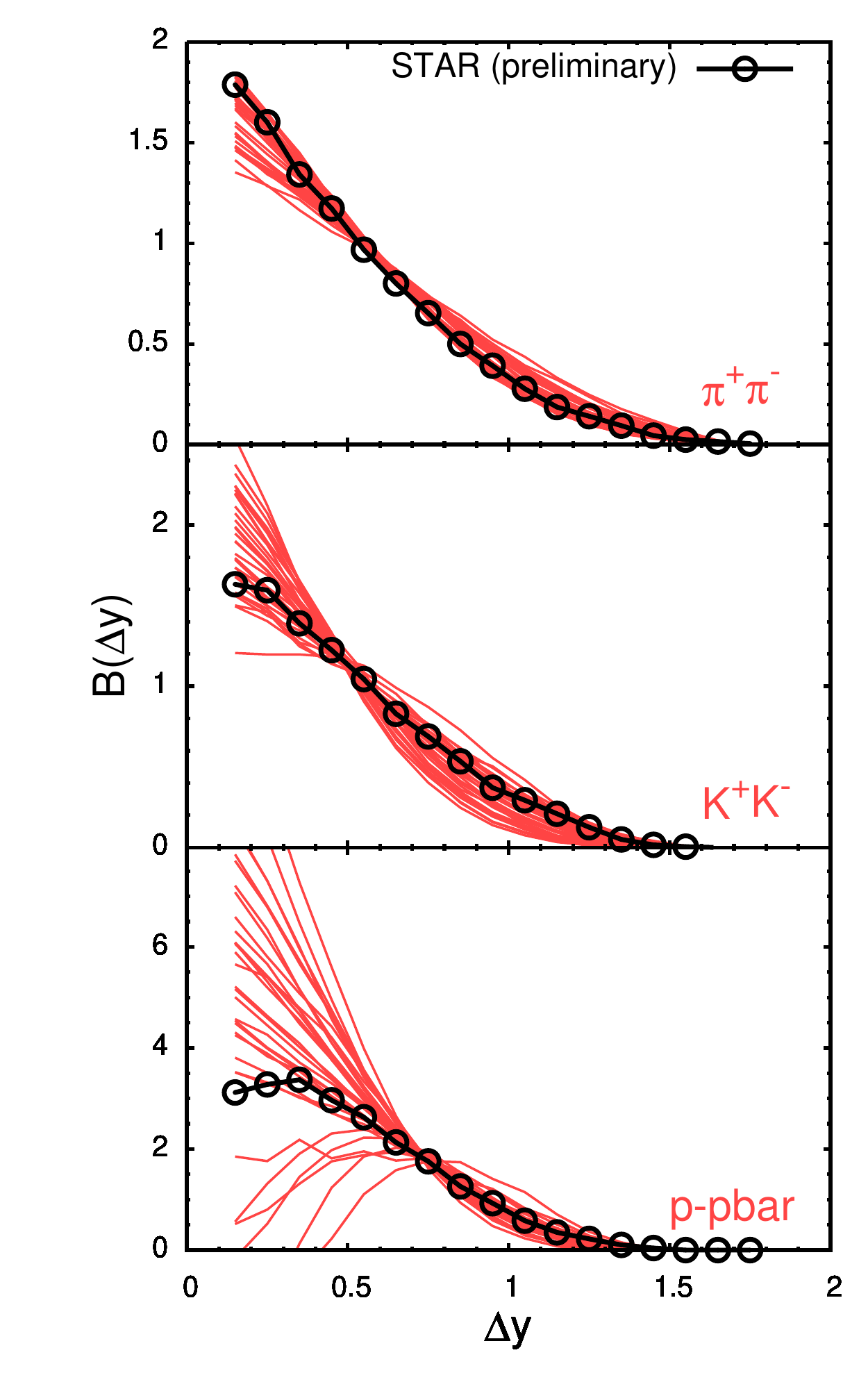}~~~
\includegraphics[width=0.44\textwidth]{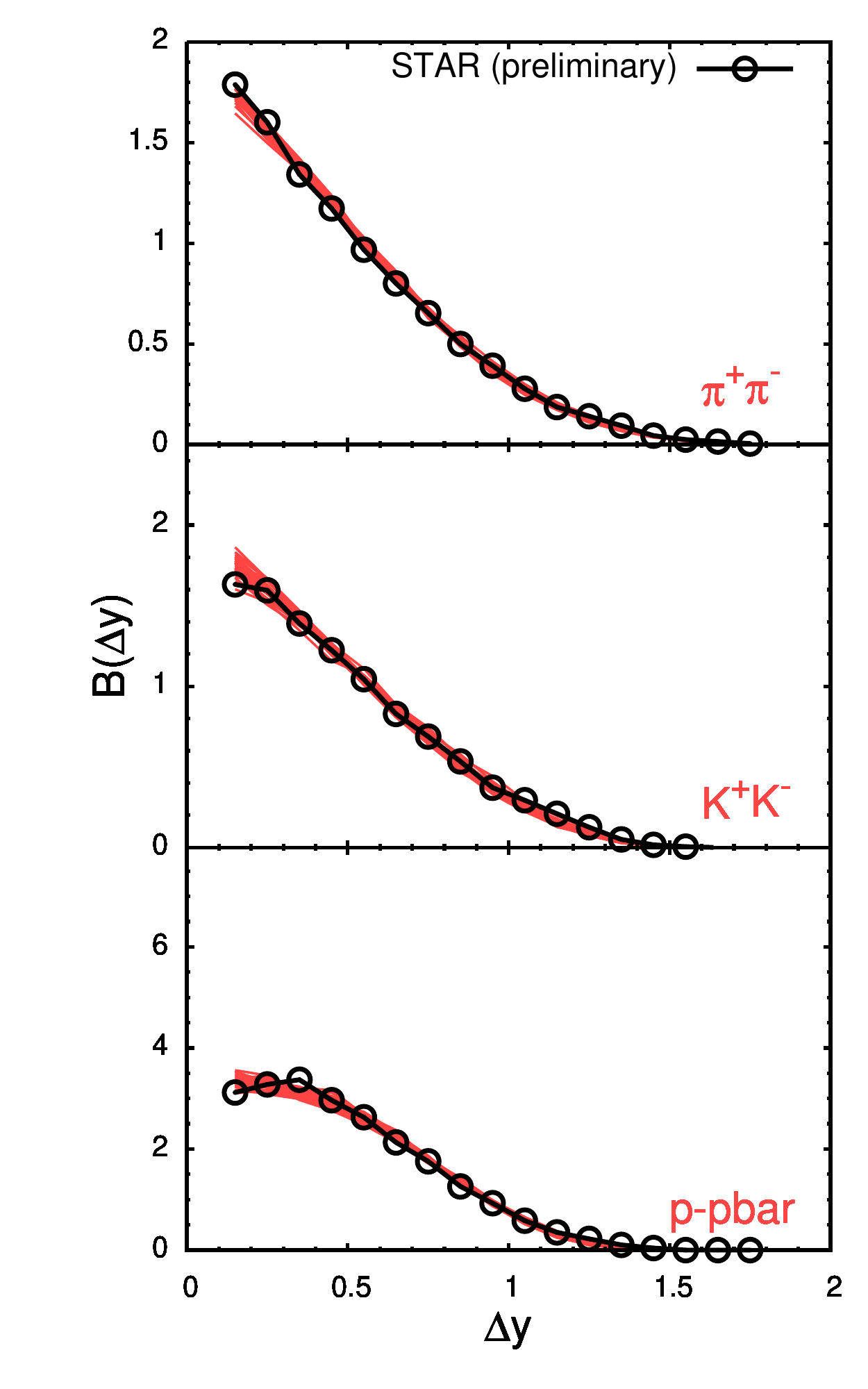}}
\caption{\label{fig:priorposterior}
Left panels: Balance functions calculated from 32 points randomly taken from the original parameter space (the prior). Calculations are shown in red for $\pi^+\pi^-$, $K^+K^-$ and $p\bar{p}$ balance functions. Right panels: Same as the left, but with 32 points taken from the MCMC trace (representative of the posterior). This illustrates the degree to which the balance functions are being fit, and also demonstrates the efficacy of the statistical procedure. In each case, the balance functions have been normalized to have a fixed area below the curves in some region (described in text) to account for the lack of knowledge about the overall efficiency. 
}
\end{figure}

Parameter space was explored through a weighted MCMC trace. The trace was weighted by a likelihood calculated by summing the $\chi^2$ for each point of the three balance functions.
\begin{equation}
\chi^2\equiv\sum_{i} (y_i-y_i^{\rm(exp)})^2 / \sigma_i^2.
\end{equation}
The likelihood was then assumed to behave $\sim e^{-\chi^2/2}$. Assigning the errors, $\sigma_i$, determines how strongly the balance functions are pushed to fit the experimental data. Errors were assigned, rather arbitrarily, as $\sigma(\Delta y)=0.1\sqrt{1.8-\Delta y}$, where $\Delta y$ is the relative rapidity used to index the balance functions. The functional form was motivated by fact that the errors must go to zero for $\Delta y$ at the maximum value measurable by the detectors. The first bin in $\Delta y$ was neglected, and due to the lack of a rigorous model of the STAR efficiency, only the shapes of the balance functions were fit, i.e. the balance functions were normalized by dividing by the area underneath each balance function. That area also neglected the first bin for the $\pi^+$ and $\pi^-$ balance functions, and neglected the first 5 bins for the $p\bar{p}$ balance function due to the rather wildly varying first few bins. Since the balance functions are of order unity, the prefactor of 0.1 would seem to suggest the errors are being over-estimated. However, there are many points in the balance functions, and cross correlations between points are neglected, and as shown in Fig. \ref{fig:priorposterior}, this error was small enough to tightly constrain the balance function to the data. Doubling the prefactor only modestly affected the widths of the posterior distributions. 

Performing the MCMC trace required sampling ten million points in the five-dimensional parameter space. Since the calculation for each point required a few minutes of CPU time, we applied a model emulator. The same software was used to emulate models from similar problems \cite{Novak:2013bqa,Gomez:2012ak}. The emulator strategy involves interpolating data from a number of sampling runs, in this case 4096 runs. The 4096 sampling runs were performed at points spread evenly through the parameter space according to Latin hyper-cube sampling. From these runs, the data was reduced to 10 principal components, which were then interpolated during the MCMC procedure by a Gaussian-process emulator. The emulator was compared to full-model runs to test its accuracy.

\begin{figure}
\centerline{\includegraphics[width=\textwidth]{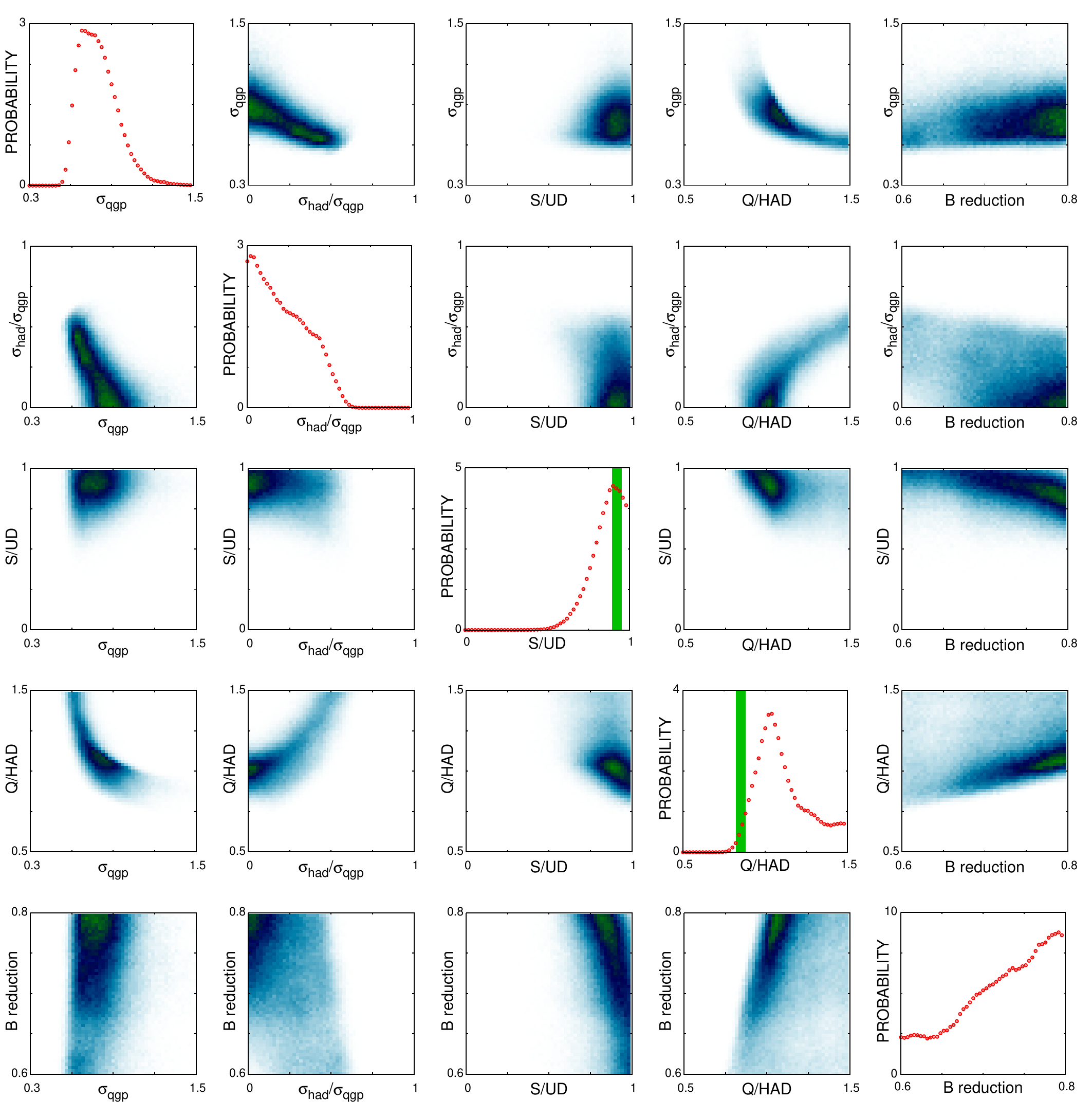}}
\caption{\label{fig:mcmc5}
Posterior distribution of parameters from MCMC trace. Along the diagonal, distributions for each parameter are shown as integrated over remaining parameters. Off-diagonal plots show correlations between pairs of parameters. The two green lines show expectations of quark chemistry from lattice gauge theory assuming an isentropic expansion that hadronized at $T=170$ MeV.}
\end{figure}

Figure \ref{fig:mcmc5} displays the posterior distribution of parameters from the MCMC trace. The plots along the diagonals show the likelihood for a given parameter projected over the remaining parameters. The width of the correlations due to the first wave of charge creation, $\sigma_{(QGP)}$, was varied in the MCMC procedure between 0.3 and 1.5. The procedure prefers a range of widths from 0.6 to 1.2. The distribution of the ratio of the two widths, $\sigma_{\rm(had)}/\sigma_{\rm(qgp)}$, is constrained by the MCMC trace to be near or below 0.5, with a preference for being near zero. This is strong validation of the two-wave picture of charge production. The best ratio of strange to up or down quarks in the QGP is found to be between 0.75 and 1.0. Lattice calculations, like those shown in Fig. \ref{fig:claudia}, suggest the number should be between 0.9 and 0.95. The ratio of the total number of quarks to the total number of hadrons in the final state was preferably near 1.1. This is about 20\% higher than expectations from lattice gauge theory. Lattice calculations provide the ratio of $\chi_{ab}$ to entropy, which can be interpreted as the number of of quarks per unit entropy. This can be compared to the ratio of hadrons to entropy assuming chemical breakup and decay beginning with a temperature of 170 MeV. This provides a ratio of the number of quarks in the QGP to the number of final-state hadrons. Thus the third and fourth plots along the diagonal suggest that the quark content of the super-hadronic matter created at RHIC is within 20\% of the expected chemistry. The final plot is for the baryon-reduction factor, $B_{\rm reduction}$. There is a modest preference for higher values, but the best way to determine the parameter would be to perform a more careful analysis of yields at RHIC. The off-diagonal plots show the correlations. For example, if one knew that $B_{\rm(reduction)}$ were smaller, e.g. near 0.65, one can see that the quark to hadron ratio would fall to a little below 1.0. 

To validate the MCMC procedure and the emulator, 32 points in the five-dimensional parameter space were taken from the posterior distribution. The model was then re-run using these points in parameter space. The resulting balance functions were then compared to the data and shown in Fig. \ref{fig:priorposterior}. The calculations appear quite successful in reproducing the experimental balance functions. In contrast 32 points were randomly drawn from the prior distribution. The resulting balance functions are also shown in Fig. \ref{fig:priorposterior}, and found to strongly differ from the experimental results. 

In summary, charge balance functions from calculations described in \cite{Pratt:2012dz} were compared to preliminary STAR results. The comparison led to a determination of the quark chemistry which was within 20\% of expectations based on assumptions of an isentropic expansion of a chemically equilibrated (according to lattice) QGP. The analysis also provided strong validation of the two-wave nature of quark production in relativistic heavy ion collisions. The theoretical model could certainly become more sophisticated and realistic. Less sudden production mechanisms could lead to non-Gaussian shapes for the two waves. If early production involves the tunneling of quark-antiquark pairs through longitudinal color fields (Schwinger mechanism), the quark-antiquark pairs might have a novel separation at birth.  Annihilation at the very tail end of the reaction could represent an even smaller scale than $\sigma_{\rm(had)}$, and might explain the narrow dip in the $p\bar{p}$ balance function at small $\Delta y$. The treatment of diffusion could incorporate time-dependent diffusion constants, and diffusion could be overlayed onto full three-dimensional hydrodynamic evolutions. Finally, the decays of hadrons through the strong interaction could modify the shape of the balance function. Tracking the relative movement of conserved charges throughout the hadronic stage could be performed with microscopic simulations, but such a calculation would be challenging.

Experimental results would be more constraining if the normalization was also fit, as opposed to only fitting the shape. This would require a detailed and accurate filter for experimental acceptance and efficiency. Finally, a broader range in rapidity would be useful. The STAR time-of-flight wall extends to $\pm 0.9$ units of pseudo-rapidity. For heavy, and thus slower, particles such as protons the rapidities tend to be approximately half the pseudo-rapidities. Thus the $p\bar{p}$ balance function was effectively measured in a smaller window. A broader acceptance experiment would provide significantly better information. Even a wider acceptance for particles only identified by their electric charge would be telling. At the LHC, both the ATLAS and CMS detectors can track charges up to several units of rapidity. Finally, correlations can be studied as a function of both relative rapidity and relative azimuthal angle. The dependence of balance functions on $\Delta\phi$ provides insight into transverse flow \cite{Bozek:2004dt}, and one would expect that since large $\Delta y$ correlations correspond to early production, that the width of the balance functions in $\Delta\phi$ would grow with increasing $\Delta y$.

The field of charge correlations is emerging from its nascent state. Balance functions of non-identified particles served as qualitative evidence for delayed hadronization \cite{Adams:2003kg}, but the conclusion carried significant caveats. This study shows how his class of observables is now providing quantitative insight into the chemical evolution of the QGP. Like femtoscopic two-particle correlations, charge balance correlations intrinsically carry six dimensions of information, and the comparison of balance functions for different species provide even richer and constraining insight. It appears that this field will soon be inundated with a flux of new experimental results from STAR, ALICE, CMS and ATLAS. This should make for an exciting time the next few years.


\end{document}